\documentclass[twocolumn]{article}

\makeatletter
\newcommand{\icmltitle}[1]{\title{#1}}
\newcommand{\icmltitlerunning}[1]{}
\newcommand{\icmlauthor}[2]{\author{#1}}
\newcommand{\icmlaffiliation}[2]{}
\newcommand{\icmlcorrespondingauthor}[2]{}
\newcommand{\icmlkeywords}[1]{\small\textbf{Keywords:} #1}
\newcommand{\printAffiliationsAndNotice}[1]{}

\makeatother

\usepackage[T1]{fontenc}
\usepackage{amsmath,amsfonts,amssymb}
\usepackage{graphicx}
\usepackage{booktabs}
\usepackage{hyperref}
\usepackage{float}
\usepackage{braket}

\icmltitle{Adaptive Control of Stochastic Error Accumulation in\\
Fault-Tolerant Quantum Computation}

\author{Tirtha Haque \\ 
\small Central University of Jammu \\ 
\small \texttt{22beece29.ece@cujammu.ac.in}}
\date{}

\begin{document}

\maketitle

\icmlkeywords{Quantum Error Correction, Reinforcement Learning, FTQC}

\begin{abstract}

In realistic hardware for quantum computation that possesses fault-tolerance, non-stationary noise and stochastic drift lead to logical failure from the temporal accumulation of errors, not from independent events. Static decoding and fixed calibration techniques are structurally incompatible with this situation because they do not take into account temporal correlations between errors or control-induced back-action of errors. These effects motivate control policies that must track noise evolution across correction cycles, rather than respond to individual syndromes in isolation.

We treat fault-tolerant quantum computation as a stochastic control problem, modelled using reduced quantum dynamics in which Pauli error processes are governed by latent noise parameters that vary temporally. From this perspective, logical failure arises through the accumulation of a hazard variable, and the corresponding control objective depends on the full history of observations.

Operating under these conditions, a Chronological Deep Q-Network (Ch-DQN) maintains an internal belief state that tracks both noise evolution and accumulated hazard. During training, backward refinement of trajectories is used to sample slowly drifting modes of operation, while runtime inference remains strictly causal. A fractional meta-update stabilizes learning in the presence of non-stationary, control-coupled dynamics.

Through multi-distance simulations that incorporate stochastic drift and feedback from decision-making, Ch-DQN suppresses hazard accumulation and extends logical survival time relative to static and recurrent baselines. Error correction in this regime is therefore no longer a static decoding task, but a control process whose success is determined over time by the underlying noise dynamics.

\end{abstract}

\section{Introduction}
\label{introduction}

  Fault-Tolerant Quantum Computation (FTQC) is the central engineering challenge required to realize scalable quantum
  processing [1, 2]. While canonical threshold theorems guarantee that arbitrarily long logical computations are
  possible below a critical physical error rate [1, 2], these guarantees typically assume bounded or weakly correlated noise models that approximate stationarity, assumptions that are increasingly strained in realistic hardware implementations [3,4].

  Practical quantum devices are subject to noise sources such as slow calibration drift [7, 8], 1/f-like long-range
  temporal correlations [5, 6], and control-induced back-action [9] that evolve on timescales comparable to or longer
  than the error-correction cycle itself. As a direct consequence, logical failure rarely stems from a single
  catastrophic event, but emerges through the gradual, long-term accumulation of correlated noise [10, 11]. This
  understanding reveals a structural mismatch with most existing approaches to Quantum Error Correction (QEC). Classical
  decoders, such as Minimum Weight Perfect Matching (MWPM) [12], and reactive threshold-based controllers treat each syndrome measurement cycle independently [13]. The structural mismatch suggests that FTQC should be reframed not as a static decoding task, but as a sequential stochastic survival control problem [14, 15]. In particular, syndrome-based decoders operate on instantaneous measurement outcomes, whereas the underlying noise parameters evolve as hidden dynamical variables, rendering the decoding problem partially observable.

  In this paper we model the drifting noise through a slowly evolving latent state, yielding a Hidden Markov Model (HMM) representation that captures effective long-range correlations in the observable syndromes where slow latent dynamics induce long temporal correlations in the observable measurement process (16).
     
\begin{figure}
    \raggedright
    \includegraphics[width=1.1\linewidth]{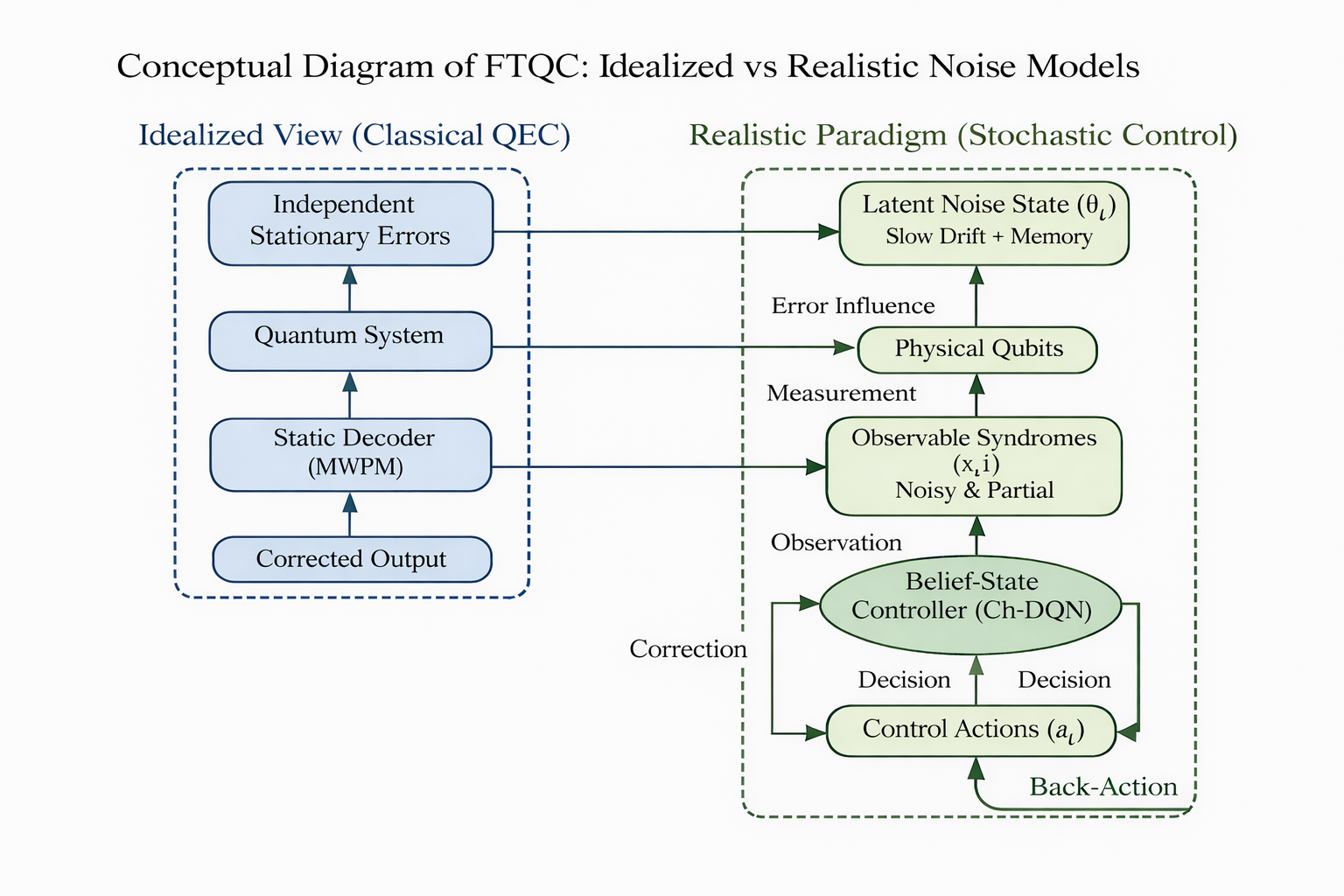}
    \caption{\protect\footnotesize
    Conceptual Representation: From Idealized to Realistic Partially Observable, Control-Coupled Dynamics}
    \label{fig:placeholder1}
\end{figure}

  This paper formalizes FTQC under drifting noise as a non-stationary, policy-coupled partially observable control problem. We introduce Chronological Deep Q-Networks (Ch-DQN), a belief-state reinforcement learning architecture that integrates causal filtering, backward latent smoothing during training, and long-memory parameter regularization. Through multi-distance simulations, we demonstrate that belief-state control suppresses hazard growth and extends logical survival relative to static and recurrent baselines.

\section{Open Quantum System Description}

\subsection{Stochastic Noise Characterization}

Fault-tolerant quantum computation is performed on physical qubits that remain continuously coupled to an uncontrolled environment [17]. The joint system–environment state,
$\rho_{SE}(t) \in \mathcal{D}(\mathcal{H}_S \otimes \mathcal{H}_E)$,
evolves under a time-dependent Hamiltonian [18],

\[
H(t)=H_S+H_E+H_{SE}(t).
\]

The interaction term is modeled as

\[
H_{SE}(t)
=
\sum_{\alpha \in \{X,Y,Z\}}
\lambda_\alpha(t)\,
\sigma_\alpha \otimes B_\alpha ,
\]

where $\sigma_\alpha$ are Pauli operators acting on system qubits and
$B_\alpha$ act on environmental degrees of freedom [19].  
In realistic quantum devices the coupling strengths $\lambda_\alpha(t)$ are not constant and often exhibit significant temporal variability caused by calibration drift, environmental fluctuations, and control back-action [18,19].

We therefore decompose the coupling into two stochastic components,

\[
\lambda_\alpha(t)
=
\bar{\lambda}_\alpha(t)
+
\zeta_\alpha(t),
\]

where $\bar{\lambda}_\alpha(t)$ captures slow non-stationary drift and
$\zeta_\alpha(t)$ captures stationary long-memory fluctuations.

\paragraph{Slow Drift Component.}

The process $\bar{\lambda}_\alpha(t)$ models effects such as thermal drift,
device aging, and quasi-static calibration offsets observed in superconducting and solid-state qubits [5,7].
Over finite calibration windows these effects are commonly approximated by a random-walk model [19]:

\[
\bar{\lambda}_\alpha(t+1)
=
\bar{\lambda}_\alpha(t)
+
\nu_\alpha(t),
\qquad
\nu_\alpha(t) \sim \mathcal{N}(0,\sigma_\nu^2),
\]

with $\mathbb{E}[\nu_\alpha(t)] = 0$.
This unit-root process is non-stationary with variance increasing linearly in time.
In practice, periodic hardware recalibration prevents indefinite divergence, so the model is understood to hold over bounded operational horizons.

\paragraph{Long-Memory Fluctuation Component.}

The process $\zeta_\alpha(t)$ models stationary correlated noise such as
$1/f$-type fluctuations widely observed in superconducting and spin-based qubits [5,6].  
We assume $\mathbb{E}[\zeta_\alpha(t)] = 0$ and

\[
\mathbb{E}[\zeta_\alpha(t)\zeta_\alpha(t+\tau)]
=
C_\alpha \tau^{-\beta},
\qquad 0 < \beta < 1,
\]

which defines a power-law autocorrelation characteristic of long-memory processes.
Such scale-invariant noise processes are commonly modeled using fractional Gaussian noise or related spectral models in qubit noise spectroscopy [5].

\paragraph{Independence Assumption.}

We assume that $\bar{\lambda}_\alpha(t)$ and $\zeta_\alpha(t)$ are independent stochastic processes:

\[
\mathbb{E}\!\left[
\bar{\lambda}_\alpha(t)\zeta_\alpha(s)
\right]
=
\mathbb{E}[\bar{\lambda}_\alpha(t)]
\mathbb{E}[\zeta_\alpha(s)]
\quad
\forall t,s.
\]

Under the zero-mean assumptions above, the covariance of
$\lambda_\alpha(t)$ decomposes additively into drift and long-memory components.

\paragraph{Resulting Noise Structure.}

The superposition of a non-stationary random walk and a stationary long-memory process produces effective coupling trajectories exhibiting both slow drift and persistent temporal correlations.
Such combined noise structures have been experimentally observed in superconducting qubits and other solid-state quantum devices [6,19].
These dynamics induce non-stationarity in the reduced system evolution over repeated QEC cycles.

\subsection{Reduced Dynamics and QEC Cycles}

Quantum error correction proceeds in discrete cycles of duration $\Delta t$ determined by the physical measurement and control latency of the hardware [22].  
Over each cycle the joint system–environment state evolves under the full Hamiltonian, and the reduced system dynamics are obtained by tracing out the environmental degrees of freedom [17].

At the cycle timescale we assume a coarse-grained description in which the system evolution is represented by a Completely Positive Trace-Preserving (CPTP) map,

\[
\mathcal{E}_t : \rho_S(t) \mapsto \rho_S(t+1).
\]

This CPTP approximation assumes that system–environment correlations are effectively refreshed at the start of each QEC cycle through measurement and control operations, a standard assumption in coarse-grained open-system models [17].

Although the underlying microscopic dynamics may contain temporal correlations, coarse-graining over a cycle produces an effective CPTP map whose parameters depend on the stochastic couplings $\lambda_\alpha(t)$ described in Section 2.1.

In stabilizer codes, repeated projection into the code space allows the noisy evolution on the logical subspace to be expressed, up to stabilizer equivalence, as a Pauli channel [11]:

\[
\mathcal{E}_t(\rho)
=
\sum_{P \in \{I,X,Y,Z\}}
p_P(t)\, P \rho P .
\]

The Pauli error probabilities $p_P(t)$ depend on the instantaneous physical parameters $\lambda_\alpha(t)$ and therefore inherit their stochastic drift and correlations.
In practice, these parameters are estimated experimentally using techniques such as randomized benchmarking and quantum process tomography [20].

Even when the latent coupling parameters evolve according to a Markov process, the observable error process need not be Markovian due to partial observability and nonlinear projection into the code space [17].

From the perspective of the classical controller:

\begin{itemize}

\item The environment is partially observed, since the full system–bath dynamics are not directly accessible [17].

\item The transition statistics are non-stationary due to drift in the coupling parameters $\lambda_\alpha(t)$ [19].

\item The observed data (syndromes) are indirect projections of an evolving stochastic process governed by hidden physical parameters.

\end{itemize}

This reduction provides the formal bridge from microscopic open-system dynamics to a high-level stochastic control model defined at the granularity of QEC cycles.

\section{Stochastic Control Formulation}

\subsection{Latent Noise State Definition}

The instantaneous Pauli error probabilities $p_P(t)$ are driven by physical coupling parameters $\lambda_\alpha(t)$, which are coupled and slowly evolving. Crucially, future error statistics are determined not by the current syndrome alone, but by the underlying physical noise regime [27,28].

We therefore define a low-dimensional latent noise state $\theta_t \in \mathbb{R}^m$ as a deterministic function of the underlying coupling parameters:
\[
\theta_t =
\begin{bmatrix}
\lambda_X^2(t) \\
\lambda_Z^2(t) \\
c_t
\end{bmatrix},
\]
where $c_t$ denotes a scalar variable capturing the effective temporal persistence of correlated fluctuations. The components of $\theta_t$ parameterize the effective bit-flip noise power, phase-flip noise power, and the strength of temporal correlations, respectively.

This vector $\theta_t$ is the true system state in the control-theoretic sense: it determines the transition statistics of future Pauli error probabilities [29]. However, $\theta_t$ is not directly observable; it can only be inferred through its effect on the syndrome statistics $x_t$. The resulting control problem is therefore partially observed.

\subsection{Latent State Dynamics}

The latent noise state $\theta_t \in \mathbb{R}^m$ captures slowly varying
hardware parameters governing the effective Pauli error channel at the
granularity of quantum error-correction (QEC) cycles.
Its evolution reflects calibration drift, environmental fluctuations,
and regime instability in the underlying physical device [30,31].

We model this evolution as a stable linear stochastic dynamical system:
\begin{equation}
\theta_{t+1} = A \theta_t + \eta_t,
\end{equation}
where $A \in \mathbb{R}^{m \times m}$ satisfies $\rho(A) < 1$,
with $\rho(\cdot)$ denoting the spectral radius, and
\[
\eta_t \sim \mathcal{N}(0, \Sigma_\eta)
\]
represents slow stochastic drift.

When $\rho(A) \approx 1$, the system exhibits long correlation times
relative to the QEC cycle duration.
Over operational horizons of interest, this near-unit-root regime
generates slowly decaying temporal correlations in the latent state.
We treat this as an effective macroscopic approximation of slow drift
at the controller timescale rather than as a microscopic derivation
of a specific spectral density (e.g., exact $1/f$ noise) [32].

Observable Pauli error probabilities are generated from the latent regime
through a simplex-constrained nonlinear mapping:
\begin{equation}
p_P(t) =
\frac{\exp\!\left(w_P^\top \theta_t\right)}
{\sum_{Q \in \{I,X,Y,Z\}} \exp\!\left(w_Q^\top \theta_t\right)}.
\end{equation}

This softmax parameterization guarantees
\[
p_P(t) \ge 0,
\qquad
\sum_{P \in \{I,X,Y,Z\}} p_P(t) = 1,
\]
ensuring that the induced channel
\begin{equation}
\mathcal{E}_t(\rho)
=
\sum_{P \in \{I,X,Y,Z\}}
p_P(t)\, P \rho P
\end{equation}
is a valid Pauli channel.

Since Pauli channels are convex combinations of unitary Kraus operators,
this construction guarantees complete positivity and trace preservation (CPTP)
for all reachable values of $\theta_t$ [33].

Although the latent process $\theta_t$ evolves Markovianly,
it is not directly observable.
The controller receives only syndrome-derived features $x_t$,
which constitute nonlinear and information-reducing projections
of the underlying noise regime.

Consequently, even if $\theta_t$ evolves according to a Markov process,
the observable error dynamics need not be Markovian in $x_t$.
From the controller’s perspective, the system therefore forms a
Partially Observable Markov Decision Process (POMDP) [34],
in which belief over $\theta_t$ serves as the sufficient statistic
for optimal control.

\subsection{Hazard Accumulation and Logical Failure}

Logical failure is formulated as a first-passage event driven by
cumulative degradation of the encoded logical state.

Let
\[
\mathcal{E}_t(\cdot \mid a_t)
\]
denote the effective logical completely positive trace-preserving (CPTP)
map applied during one QEC cycle under latent regime $\theta_t$
and control action $a_t$.
The cumulative logical channel after $t$ cycles is given by
\begin{equation}
\mathcal{E}_{0:t}
=
\mathcal{E}_t
\circ
\mathcal{E}_{t-1}
\circ
\cdots
\circ
\mathcal{E}_0 .
\end{equation}

For a fixed encoded logical state $\ket{\psi_L}$,
the logical fidelity at cycle $t$ is defined as
\begin{equation}
F_L(t)
=
\bra{\psi_L}
\mathcal{E}_{0:t}
\!\left(
\ket{\psi_L}\bra{\psi_L}
\right)
\ket{\psi_L}.
\end{equation}

We define the instantaneous logical risk as the one-step
decrement in fidelity:
\begin{equation}
\rho(\theta_t, x_t, a_t)
:=
F_L(t) - F_L(t+1),
\qquad
\rho \ge 0.
\end{equation}

The cumulative hazard process is therefore
\begin{equation}
H_t
:=
\sum_{\tau=0}^{t-1}
\rho(\theta_\tau, x_\tau, a_\tau)
=
1 - F_L(t).
\end{equation}

Hence, the hazard variable coincides exactly with the
accumulated logical infidelity relative to the initial encoded state.

Logical failure is defined as the first passage time of the hazard
process across a code-dependent tolerance threshold:
\begin{equation}
T_{\mathrm{fail}}
=
\inf
\left\{
t :
H_t \ge H_{\mathrm{crit}}(d)
\right\}.
\end{equation}

The threshold $H_{\mathrm{crit}}(d)$ reflects the logical tolerance
of a surface code of distance $d$ and increases with code size [35].

The control objective is to maximize the survival functional
\begin{equation}
\mathbb{E}[T_{\mathrm{fail}}],
\end{equation}
which depends on the entire control trajectory through its
influence on the latent regime $\theta_t$ and,
consequently, on the long-horizon rate of logical fidelity decay.

Unlike myopic decoders that minimize instantaneous physical
error probability, this formulation treats quantum error correction
as a long-horizon survival control problem.

\subsection{Observability Limits and the Need for Belief States}

The latent noise state $\theta_t$ introduced in Sections 3.1–3.2 is never directly observable in a real FTQC system. At each cycle, the controller observes only a low-dimensional projection of the hardware state extracted from stabilizer measurements and derived features:
\[
x_t = (\rho_t, \sigma_t, \pi_t, H_t).
\]

These quantities are noisy, indirect functions of the underlying latent regime $\theta_t$.

Even if the latent dynamics $\theta_{t+1} = A\theta_t + \eta_t$ are Markovian, the observable process $\{x_t\}$ is not generally Markov.

Thus the system forms a partially observable Markov decision process (POMDP) [36]. The sufficient statistic for control is the posterior belief over the latent noise regime:
\[
b_t(\theta)
=
p(\theta_t = \theta \mid x_{0:t}, a_{0:t-1}).
\]

An optimal control policy must therefore be a function of the belief state:
\[
a_t = \pi(b_t).
\]

Ch-DQN addresses this limitation by learning a finite-dimensional embedding
\[
h_t \approx \Pi(b_t)
\]
that approximates the sufficient statistic required for belief-state control.
\section{Control--Noise Back-Action}

\subsection{Policy-Coupled Transitions}

In fault-tolerant quantum hardware, control is not passive. 
Interventions intended to suppress current errors perturb the physical device and may alter its subsequent noise statistics 
[21,25]. 
This establishes a feedback loop:
\[
\text{noise} \;\xrightarrow{\text{syndrome}}\; \text{control} \;\xrightarrow{\text{back-action}}\; \text{future noise}.
\]

To capture this effect, we extend the latent dynamics to include action dependence:
\[
\theta_{t+1} = A\theta_t + B a_t + \eta_t,
\]
where $a_t$ is the applied control action and $B$ encodes how interventions modify the latent noise regime.

This formulation does not imply that the environment transition law depends on the learned policy itself; rather, it reflects that the physical dynamics depend on the realized control actions 
[14,35]. 
The transition kernel
\[
p(\theta_{t+1} \mid \theta_t, a_t)
\]
is stationary in parameters $(A,B,\Sigma_\eta)$ but control-dependent in state evolution.

Actions such as recalibration pulses may reduce instantaneous error rates while shifting bias points or injecting additional disturbance, creating a bias–variance tradeoff in the effective noise process [18,19].

\begin{figure}
    \centering
    \includegraphics[width=1\linewidth]{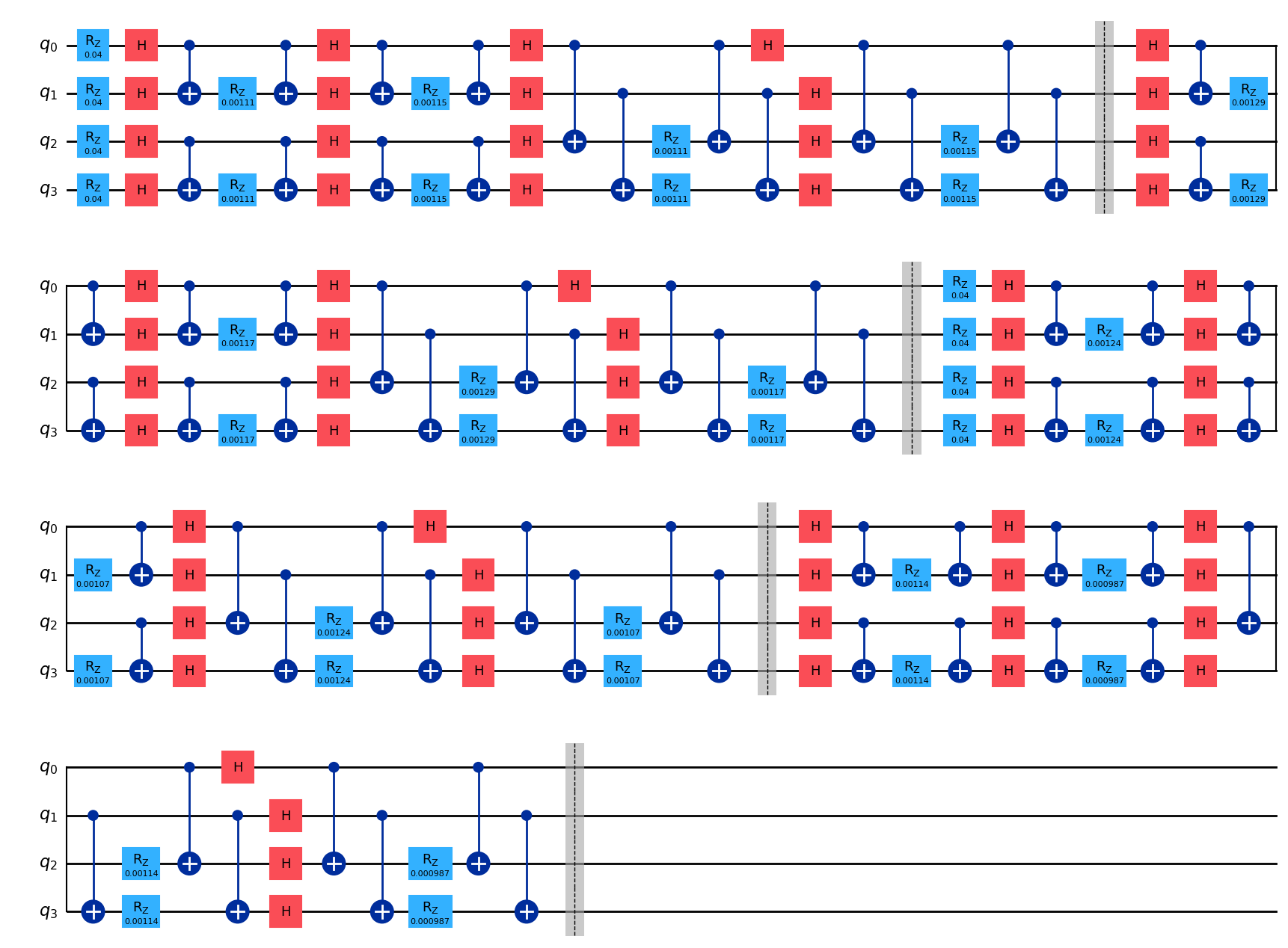}
    \caption{\protect\footnotesize Example realization of control-induced Hamiltonian modulation. Control pulses applied to system qubits modify subsequent XX and ZZ interactions}
    \label{fig:placeholder2}
\end{figure}

\subsection{Non-Stationarity and Survival Objective}

Two distinct sources of non-stationarity arise in this setting.

First, the latent noise regime drifts autonomously:
\[
\theta_{t+1} = A\theta_t + \eta_t.
\]

Second, control modifies the latent evolution through $B a_t$, introducing closed-loop coupling between estimation and intervention [14].

Although the latent process is Markovian, the observable process remains partially observed and non-stationary in observation space [29,34].

The objective is to delay logical failure, defined as the first-passage time of the hazard process,
\[
T_{\text{fail}} = \inf\{t : H_t \ge H_{\text{crit}}(d)\}.
\]

We therefore consider the long-horizon survival objective:
\[
\max_{\pi} 
\;
\mathbb{E}_\pi
\!\left[
T_{\text{fail}} 
-
\alpha 
\sum_{t < T_{\text{fail}}} \|a_t\|
\right],
\]
where the expectation is taken over trajectories induced by the belief-state policy $\pi(b_t)$.

This objective is Markov in belief space but not in observation space, reinforcing the necessity of belief-state control [27,29,34].

\section{Failure of Memoryless Policies}

The discussed control problem is formally a Partially Observable Markov Decision Process [29,30]. 
The latent physical state $\theta_t$ is not directly observed; instead, the controller receives stabilizer-derived features $x_t$, which constitute a noisy and lossy projection of the underlying hardware dynamics.

Consequently, the observation process $\{x_t\}$ is not Markovian in general, even if the latent process $\{\theta_t\}$ is. In particular,
\[
p(x_{t+1} \mid x_t, a_t)
\neq
p(x_{t+1} \mid x_{0:t}, a_{0:t}),
\]
whenever the latent state is not fully revealed.

Memoryless control policies—including static decoders, fixed threshold rules, and feedforward RL agents—implicitly assume that $x_t$ is a sufficient statistic. Under drift and partial observability, this assumption is invalid, and such policies cannot in general achieve optimal long-horizon survival [27,30].

\subsection{Belief State as a Sufficient Statistic}

Because $\theta_t$ is unobserved, optimal control must operate on the posterior belief over the latent regime,
\[
b_t(\theta)
=
p(\theta_t = \theta \mid x_{0:t}, a_{0:t-1}).
\]

The belief evolves according to the Bayesian filtering equation [23]:
\[
b_{t+1}(\theta')
=
\frac{
p(x_{t+1} \mid \theta')
\int
p(\theta' \mid \theta, a_t)\,
b_t(\theta)\, d\theta
}{
p(x_{t+1} \mid x_{0:t}, a_{0:t})
}.
\]

This belief state is a sufficient statistic for control: there exists an optimal policy of the form
\[
a_t = \pi(b_t),
\]
and the resulting decision process is Markov in belief space [29,34].

Ch-DQN does not represent the full posterior distribution. Instead, it learns a low-dimensional embedding $h_t$ that approximates a summary of $b_t$, for example its conditional mean,
\[
h_t \approx \mathbb{E}[\theta_t \mid x_{0:t}, a_{0:t-1}],
\]
providing a compressed, approximately Markovian representation of the system history.

Classical estimators fail due to model uncertainty, parameter drift, policy back-action, and fractional or long-term memory [17,19]. Memory-based RL controllers overcome these by learning a low-dimensional belief summary $h_t$ that approximates the sufficient statistic $b_t$. By capturing non-local dependencies within a latent space, these controllers enable optimal control over non-stationary noise regimes where Markovian filters are suboptimal, providing a sufficient, approximately Markovian statistic for the unobserved system history [27,33].

\begin{figure}
    \centering
    \includegraphics[width=1\linewidth]{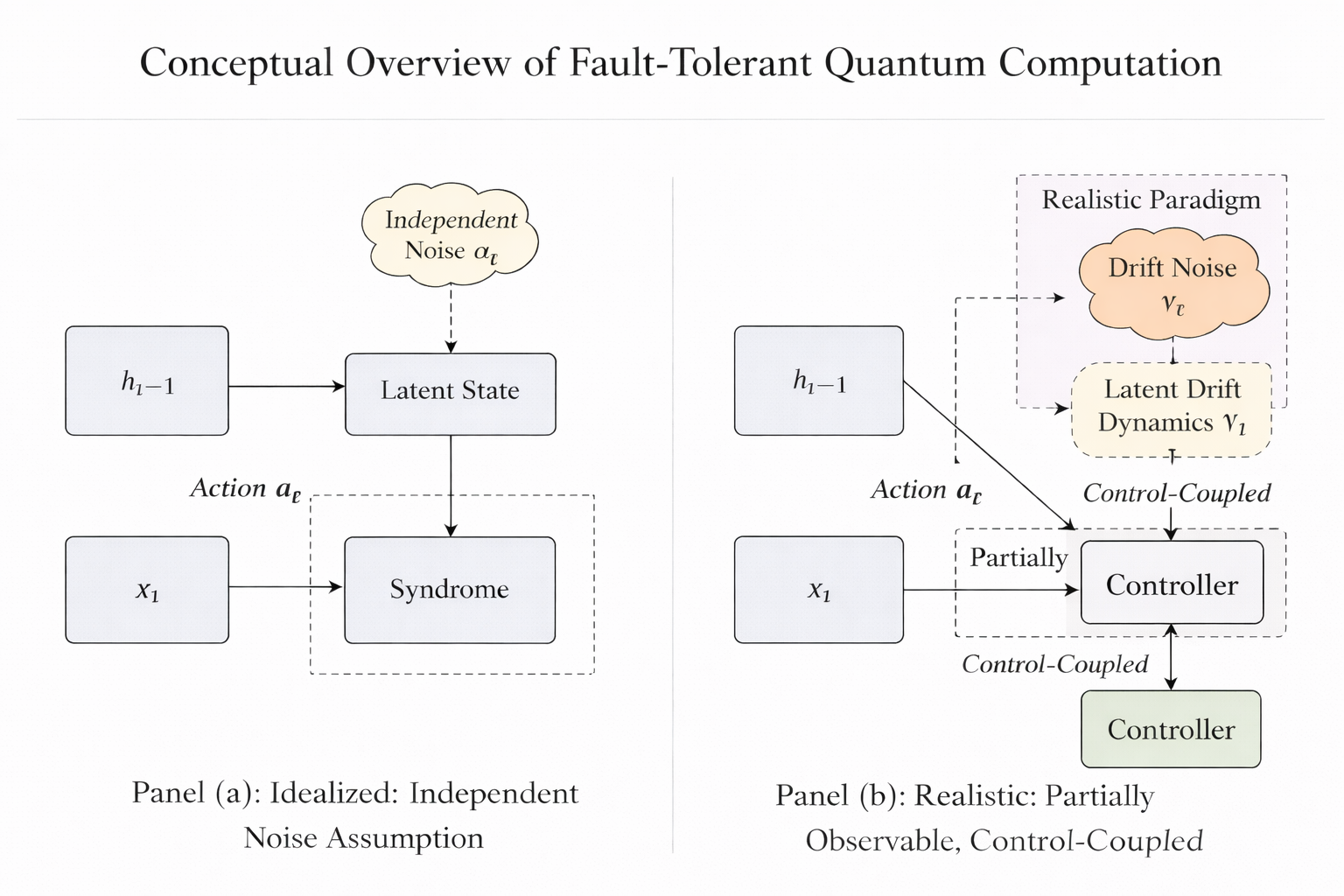}
    \caption{\protect\footnotesize Belief-based control under drift versus independent noise model}
    \label{fig:controller}
\end{figure}
\section{Chronological Deep Q-Learning (Ch-DQN)}

The partially observable structure of FTQC under drifting noise implies that the observation sequence $x_t$ is generally non-Markovian, even when the latent physical state $\theta_t$ evolves Markovianly [29,30]. To enable value-based reinforcement learning, we introduce \textbf{Chronological Deep Q-Learning (Ch-DQN)}, an architecture that learns a compact latent representation $h_t$ designed to summarize the relevant history for long-horizon survival control.

Rather than operating directly on insufficient observations $x_t$, Ch-DQN learns a latent embedding $h_t$ intended to approximate a summary of the posterior belief over the unobserved noise regime [33,34]. This learned representation allows the value function to be defined over an approximately Markovian latent space.
.

\subsection{Belief-State Representation}

The FTQC environment is governed by the latent state $\theta_t$ and the policy-coupled transition dynamics described previously. 

At each QEC cycle the controller observes a low-dimensional feature vector
\[
x_t = (\rho_t, \sigma_t, \pi_t, H_t) \in \mathbb{R}^4,
\]
where $\rho_t$ represents an instantaneous logical risk signal, $\sigma_t$ is a noise proxy derived from the current operating regime, $\pi_t$ is a threshold indicator capturing whether the noise proxy exceeds a safety margin, and $H_t$ denotes the accumulated hazard variable defined in Section~3.3. These quantities constitute a noisy and partial projection of the underlying latent noise regime $\theta_t$ [5,17,19].

The controller selects a discrete control action
\[
a_t \in \mathcal{A}, \quad \mathcal{A} = \{0,1,2\},
\]
where $a_t=0$ corresponds to no corrective intervention and larger values represent increasing corrective pulse strength applied to the system. Actions influence both the immediate reward and the subsequent evolution of the latent noise regime through the policy-coupled dynamics described in Section~4 [14].

The sufficient statistic for optimal control is the posterior belief
\[
b_t(\theta) = p(\theta_t = \theta \mid x_{0:t}, a_{0:t-1}),
\]
which evolves according to the Bayesian filtering recursion [23].

Exact filtering is computationally intractable for this nonlinear, non-stationary system [33]. We therefore approximate the belief state using a learned, low-dimensional embedding $h_t \in \mathbb{R}^m$:
\[
h_t \approx \Phi(b_t),
\]
where $\Phi$ denotes a learned compression of the posterior distribution. In practice, $h_t$ may approximate summary statistics such as the conditional mean,
\[
h_t \approx \mathbb{E}[\theta_t \mid x_{0:t}, a_{0:t-1}],
\]
though no parametric assumption on the posterior is imposed.

The objective is not to recover $\theta_t$ exactly, but to learn a representation sufficient for accurate long-horizon value estimation [27].

\subsection{Forward Latent Filtering}

At deployment time, decisions must be causal. The latent representation is updated sequentially via a nonlinear recurrent mapping,
\[
h_t = f_\psi(h_{t-1}, x_t),
\]
where $f_\psi$ is parameterized as
\[
h_t = \tanh(W h_{t-1} + V x_t + R r_t).
\]

This recurrence functions as a learned causal filter, analogous in structure to the update step of classical Bayesian filters [23]. The parameters $(W,V,R)$ are trained end-to-end so that $h_t$ captures predictive information about future survival risk.

Importantly, this filtering mechanism does not assume linear-Gaussian dynamics; it learns the effective belief update directly from interaction data [27,41].

\subsection{Backward Latent Refinement (Smoothing)}

While deployment must be causal, training can leverage full trajectories. In temporally correlated systems, future observations provide information about earlier latent states [23]. 

To exploit this structure during training, we introduce a bidirectional refinement step:
\[
\tilde{h}_t = g_\psi(h_{t-1}, h_{t+1}, x_t),
\]
implemented as
\[
\tilde{h}_t =
\tanh(W h_{t-1} + U h_{t+1} + V x_t + R r_t).
\]

This smoothing step incorporates future context to improve the internal representation used for value estimation. The refined latent $\tilde{h}_t$ is used only during training and does not alter the causal deployment architecture.

\begin{figure}
    \centering
    \includegraphics[width=1\linewidth]{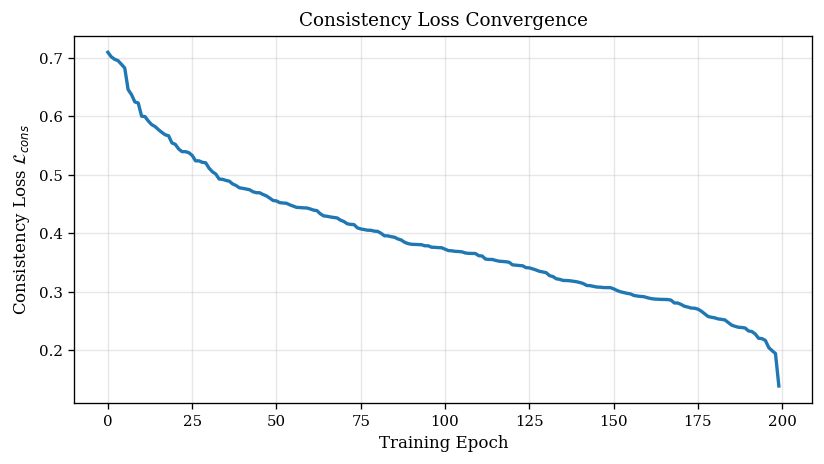}
    \caption{\protect\footnotesize loss convergence in Ch-DQN}
    \label{fig:loss}
\end{figure}

\subsection{Fixed-Point Consistency Loss}

The forward latent filter operates causally and therefore does not have access to future observations. During training, however, the bidirectional refinement step produces a smoothed representation $\tilde{h}_t$ that incorporates information from the full trajectory.

The controller is trained using a reward signal that balances risk suppression and control cost. At each time step the environment provides the scalar reward
\[
r_t = -\rho_t - \lambda_a a_t ,
\]
where $\rho_t$ is the instantaneous logical risk signal and $a_t$ denotes the applied control action. The coefficient $\lambda_a$ penalizes excessive interventions and encourages parsimonious control policies. This reward structure encourages the agent to suppress hazard accumulation while minimizing unnecessary corrective pulses [27].

To align the causal representation with its smoothed counterpart, we introduce a fixed-point consistency loss:
\[
\mathcal{L}_{\mathrm{cons}} =
\sum_{t=0}^{T}
\left\|
h_t^{\mathrm{online}}
-
\tilde{h}_t
\right\|^2.
\]

This loss encourages the causal filter to approximate the representation that would be obtained under fixed-interval smoothing. The refined latent state $\tilde{h}_t$ is used only during training; inference remains strictly causal.

This mechanism may be interpreted as temporal knowledge distillation, where a future-informed teacher representation guides the causal student network toward improved long-horizon predictive structure without introducing acausality at deployment [33,34].

\section{Meta-Stability and Learning Under Drift}

The FTQC control problem induces a non-stationary, policy-coupled stochastic process. The latent noise regime evolves as
\[
\theta_{t+1} = A\theta_t + B a_t + \eta_t,
\qquad
\eta_t \sim \mathcal{N}(0,\Sigma_\eta),
\]
with $\rho(A) \lesssim 1$ [5,17,19]. The transition kernel therefore depends on the control policy $\pi$ through $a_t = \pi(h_t)$, implying that the environment transition law
\[
P_\pi(\theta_{t+1} \mid \theta_t)
\]
changes as the policy parameters evolve.

Let $d_t^\pi(s)$ denote the state distribution induced by policy $\pi$ at iteration $t$. Because $\pi$ is updated during learning, we obtain a sequence of distributions
\[
d_t^{\pi_t} \neq d_t^{\pi_{t+1}},
\]
even under identical initial conditions. This distributional drift violates the stationary sampling assumption underlying classical Q-learning convergence analyses [27,41].

For long-horizon survival objectives, the gradient of the expected return
\[
J(\pi) = \mathbb{E}_\pi[T_{\text{fail}}]
\]
requires backpropagation through trajectories of length $T \gg 1$. The gradient can be expressed formally as
\[
\nabla_w J
=
\mathbb{E}
\left[
\sum_{t=0}^{T}
\nabla_w \log \pi_w(a_t \mid h_t)
\, G_t
\right],
\]
where $G_t$ denotes the return-to-go [27]. In practice, truncated backpropagation replaces this with a finite-horizon approximation of length $L \ll T$, inducing bias when the underlying temporal dependence exceeds $L$ [27,33].

\paragraph{Meta-Parameter Update.}
To stabilize learning under this non-stationary regime, the controller incorporates a slow meta-update operating on historical parameter snapshots. Let $w_t$ denote the network parameters. The meta-update applies a fractional-order correction
\[
\Delta^\gamma w_t = \sum_{k=0}^{K-1} \alpha_k (w_{t-k}-w_{t-k-1}),
\]
with memory length $K$, decay exponent $\gamma$, and meta step size $\eta_{\mathrm{meta}}$.

\subsection{Recurrent Instability Under Drift}

Recurrent models parameterized by $w$ propagate hidden states according to
\[
h_t = f_w(h_{t-1}, x_t).
\]
The Jacobian of the recurrent mapping over $k$ steps is
\[
\frac{\partial h_t}{\partial h_{t-k}}
=
\prod_{j=0}^{k-1}
\frac{\partial f_w(h_{t-j},x_{t-j})}{\partial h_{t-j-1}}.
\]
Even when $\| \partial f_w / \partial h \| < 1$ locally, truncation at horizon $L$ effectively assumes negligible influence for $k > L$ [27]. When the latent process exhibits slow drift, i.e.
\[
\mathrm{Cov}(\theta_t,\theta_{t+k}) \not\approx 0
\quad \text{for large } k,
\]
this truncation induces systematic representation error [17,19].

\begin{figure}
    \centering
    \includegraphics[width=1\linewidth]{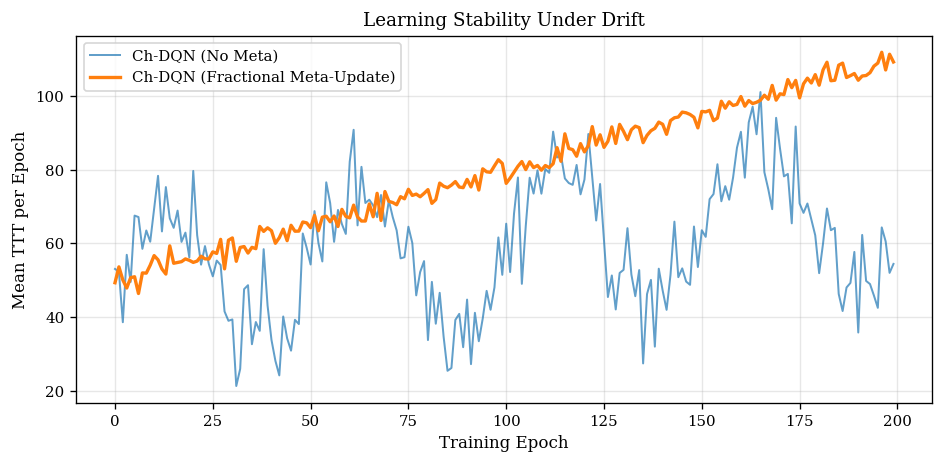}
    \caption{\protect\footnotesize Meta Layer Analysis}
    \label{fig:placeholder3}
\end{figure}

Furthermore, because $P_{\pi_t} \neq P_{\pi_{t+1}}$, recurrent representations trained under $d_t^{\pi_t}$ may become suboptimal under $d_t^{\pi_{t+1}}$, leading to instability commonly referred to as catastrophic forgetting [27,33]. This instability arises from distribution shift rather than representational incapacity.

\subsection{Fractional Meta-Update}

To stabilize optimization under distributional drift, we introduce a meta-level parameter update incorporating long-memory effects across training iterations.

Let $w_t$ denote the parameter vector after iteration $t$. Standard stochastic gradient descent performs
\[
w_{t+1} = w_t - \eta \nabla \mathcal{L}_t(w_t)
\]
[27].

To align the optimizer with the long-range correlations of the physical noise, we define a fractional-style difference operator. Unlike Markovian SGD, this discrete, non-local memory kernel enables the optimizer to track the non-stationary "drift-manifold" by incorporating the system's temporal history directly into the parameter update.

\begin{align}
\Delta^\gamma w_t
&=
\sum_{k=0}^{K-1}
\alpha_k
\left(
w_{t-k} - w_{t-k-1}
\right), \\
\alpha_k
&=
\frac{(k+1)^{-\gamma}}{\sum_{j=1}^{K} j^{-\gamma}},
\quad 0 < \gamma < 1.
\end{align}

The parameter recursion becomes
\[
w_{t+1}
=
w_t
-
\eta \nabla \mathcal{L}_t(w_t)
-
\eta_m \Delta^\gamma w_t.
\]

The additional term introduces a non-local regularization in parameter space. Under mild boundedness assumptions on $\{\nabla \mathcal{L}_t\}$ and finite $K$, the recursion can be interpreted as a discrete approximation to a fractional-order stochastic differential equation
\[
D_t^\gamma w_t = - \nabla \mathcal{L}_t(w_t) + \xi_t,
\]
where $D_t^\gamma$ denotes a fractional derivative operator and $\xi_t$ aggregates stochastic gradient noise.

Unlike exponential momentum
\[
m_t = \beta m_{t-1} + (1-\beta)\nabla \mathcal{L}_t,
\]
which induces exponentially decaying memory, the power-law kernel $\alpha_k$ decays polynomially, retaining influence from persistent parameter trends across multiple epochs [27,41].

\subsection{Stability Interpretation}

Consider the linearized dynamics around a local equilibrium $w^\star$:
\[
w_{t+1} - w^\star
=
\left(I - \eta H_t\right)(w_t - w^\star)
-
\eta_m \Delta^\gamma w_t,
\]
where $H_t$ denotes the local Hessian.

The fractional term acts as a history-dependent damping operator. When drift induces low-frequency oscillations in $w_t$, the polynomial weighting suppresses sustained deviations while preserving responsiveness to gradual shifts.

We treat this fractional update as a stability-oriented optimization heuristic motivated by long-memory stochastic approximation rather than as a physically derived necessity [17,23].

\begin{figure}
    \centering
    \includegraphics[width=1\linewidth]{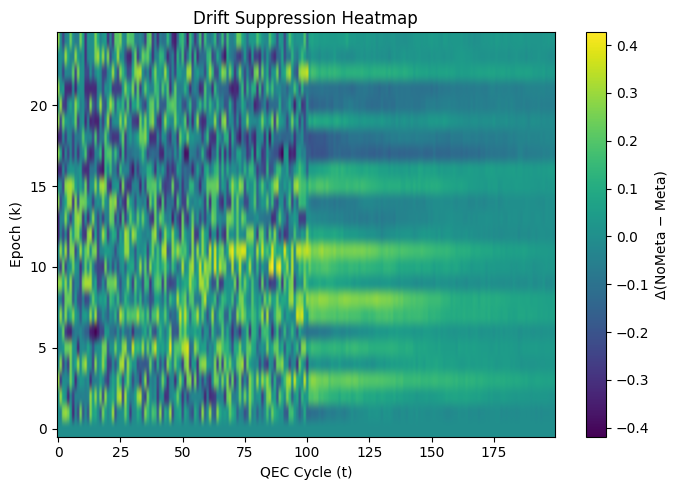}
    \caption{\protect\footnotesize Multi-panel analysis of meta-update--induced suppression of slow-time latent drift.}
    \label{fig:latent}
\end{figure}

\subsection{Two-Timescale Structure}

The combined dynamics exhibit an implicit two-timescale recursion:
\[
\begin{aligned}
h_{t+1} &= f_{w_t}(h_t,x_t), \\
w_{t+1} &= w_t - \eta \nabla \mathcal{L}_t(w_t) - \eta_m \Delta^\gamma w_t.
\end{aligned}
\]

For $\eta_m \ll \eta$, the latent representation adapts rapidly relative to the slower fractional meta-dynamics. This separation parallels classical two-timescale stochastic approximation schemes, where fast variables track local equilibria while slow variables compensate for non-stationary drift [27,28].

\subsection{Hardware Latency and Computational Scaling}

Real-time quantum error correction operates under strict timing constraints [22]. Let $T_{\text{cycle}}$ denote the duration of a single QEC cycle and $T_{\text{inf}}$ the controller inference latency. We define the latency ratio
\[
\kappa = \frac{T_{\text{inf}}}{T_{\text{cycle}}}.
\]

Standard RL formulations assume $T_{\text{inf}} \ll T_{\text{cycle}}$ [27]. In superconducting systems, $T_{\text{cycle}} \approx 1\,\mu\text{s}$, making regimes with $\kappa \gtrsim 1$ practically relevant [4,22].

\paragraph{Delayed-Feedback Regime.}

When $\kappa > 1$, actions computed at time $t$ are applied at $t+\kappa$:
\[
a_t = \pi(h_t), \quad \text{applied at } t+\kappa.
\]

Reactive decoders relying solely on $x_t$ implicitly assume zero delay [11]. If reaction time exceeds coherence time, decisions are based on outdated noise information.

\paragraph{Predictive Belief Mitigation.}

Ch-DQN maintains a belief state
\[
h_t \approx \mathbb{E}[\theta_t \mid x_{0:t}, a_{0:t-1}],
\]
where the latent regime evolves as
\[
\theta_{t+1} = A\theta_t + B a_t + \eta_t,
\qquad \rho(A) \lesssim 1
\]
[5,19].

If the correlation time $\tau_{\text{corr}}$ satisfies
\[
\frac{\kappa}{\tau_{\text{corr}}} \ll 1,
\]
then $\mathbb{E}[\theta_{t+\kappa} \mid h_t]$ remains predictable, allowing anticipatory control [14]. When $\kappa \gtrsim \tau_{\text{corr}}$, predictive advantage vanishes.

\paragraph{Inference Complexity.}

Let $d_x$ be the observation dimension, $d_h$ the latent dimension, and $K$ the action space size. The dominant forward-pass complexity per QEC cycle is
\[
\mathcal{O}(d_h^2 + d_h d_x + d_h K),
\]
which reduces to $\mathcal{O}(d_h^2)$ when $d_h \gg d_x, K$.

Thus,
\[
T_{\text{inf}} \propto d_h^2.
\]

\paragraph{Hardware Feasibility Bound.}

To ensure $\kappa \le 1$, we require
\[
c\, d_h^2 \le T_{\text{cycle}},
\]
where $c$ is a hardware-dependent constant. This implies
\[
d_h \le \sqrt{\frac{T_{\text{cycle}}}{c}},
\]
establishing a quadratic scaling limit on latent dimension.

\paragraph{Experimental Assumption.}

All experiments in Section~8 assume $\kappa \le 1$, corresponding to near-real-time inference. Delayed-control evaluation ($\kappa > 1$) is left for future work.

\section{Experimental Evaluation}

\subsection{Environment}
\label{sec:environment}

We evaluate the proposed belief-state controller on a controlled family of
non-stationary fault-tolerant quantum memories exhibiting slow drift,
temporally correlated fluctuations, and control-induced back-action
[4,5,17,19].

We simulate rotated surface codes with code distance
\[
d \in \{3,5,7\},
\]
and operate at the granularity of discrete quantum error-correction (QEC) cycles
[10,11,31,32].

% -------------------------------------------------------
\paragraph{Logical Channel Simulation.}

At each QEC cycle, physical noise induces an effective logical
completely positive trace-preserving (CPTP) map over the encoded subspace
[1,24].

Rather than employing an abstract hazard proxy,
we explicitly simulate logical state evolution under the composed channel

\[
\mathcal{E}_{0:t}
=
\mathcal{E}_t
\circ
\mathcal{E}_{t-1}
\circ
\cdots
\circ
\mathcal{E}_0.
\]

For a fixed encoded logical state $|\psi_L\rangle$,
the logical fidelity is computed as

\[
F_L(t)
=
\langle \psi_L
|
\mathcal{E}_{0:t}
\big(
|\psi_L\rangle\langle\psi_L|
\big)
|
\psi_L
\rangle.
\]

Hazard is defined directly as accumulated logical infidelity

\[
H_t = 1 - F_L(t).
\]

Thus, the simulated hazard process coincides exactly with
logical degradation of the encoded state and is not a surrogate risk measure
[1,11].

Logical failure is defined as the first-passage time

\[
T_{\mathrm{fail}}
=
\inf
\left\{
t :
H_t \ge H_{\mathrm{crit}}(d)
\right\}.
\]

% -------------------------------------------------------
\paragraph{Hazard Threshold Scaling.}

We parameterize the failure threshold as

\[
H_{\mathrm{crit}}(d) = c \sqrt{d}.
\]

This sublinear scaling reflects increased logical tolerance with
larger code distance while preserving finite-time failure under persistent drift
[11,32].

The $\sqrt{d}$ scaling avoids trivial exponential separation across distances
while maintaining comparable dynamic ranges for controller evaluation.

The constant $c$ is fixed across experiments.

% -------------------------------------------------------
\paragraph{Physical Noise Model.}

Each physical qubit is subject to a time-dependent Pauli channel
with stochastic couplings

\[
\lambda_\alpha(t)
=
\bar{\lambda}_\alpha(t)
+
\zeta_\alpha(t).
\]

The slow component follows a random walk

\[
\bar{\lambda}_\alpha(t+1)
=
\bar{\lambda}_\alpha(t)
+
\nu_\alpha(t),
\]

where

\[
\nu_\alpha(t)
\sim
\mathcal{N}(0, \sigma_\nu^2).
\]

The fast component exhibits stationary long-range temporal correlations
with power-law decay

\[
\mathrm{Cov}\!\left(
\zeta_\alpha(t),
\zeta_\alpha(t+\tau)
\right)
\propto
\tau^{-\beta},
\qquad
0 < \beta < 1
\]

consistent with experimentally observed $1/f$ noise in superconducting qubits
[5,6,7,19,26].

The combination of slow drift and correlated fluctuations
induces non-stationary effective Pauli error probabilities
at the logical level.

% -------------------------------------------------------
\paragraph{Policy-Coupled Latent Dynamics.}

Control actions influence subsequent noise evolution through
linear coupling in the latent regime

\[
\theta_{t+1}
=
A \theta_t
+
B a_t
+
\eta_t
\]

reflecting experimentally observed control–environment back-action
in quantum hardware
[14,36].

Here,
$A$ captures autonomous drift,
$B$ models control back-action,
and

\[
\eta_t \sim \mathcal{N}(0, \Sigma_\eta)
\]

represents residual stochastic perturbations.

The transition kernel

\[
p(\theta_{t+1} \mid \theta_t, a_t)
\]

is stationary in parameters $(A,B,\Sigma_\eta)$
but action-dependent in state evolution,
producing closed-loop non-stationarity during learning.

% -------------------------------------------------------

\subsection{Evaluation Metrics}

Each policy is evaluated over 500 independent Monte Carlo runs,
a standard methodology for stochastic policy evaluation
[27,28].

We report the following metrics:

\begin{itemize}

\item \textbf{TTT (Time-to-Threshold)}

\[
\text{TTT} = \mathbb{E}[T_{\text{fail}}]
\]

\item \textbf{HZ (Hazard Rate)}

\[
\text{HZ}
=
\mathbb{E}
\left[
\frac{H_{T_{\text{fail}}}}{T_{\text{fail}}}
\right]
\]

\item \textbf{CTRL (Control Cost)}

\[
\text{CTRL}
=
\mathbb{E}
\left[
\sum_{t < T_{\text{fail}}}
\|a_t\|
\right]
\]

\item \textbf{LAT\_NORM}

\[
\text{LAT\_NORM}
=
\mathbb{E}
[\|h_t\|]
\]

\end{itemize}

We additionally report empirical standard deviations and 95\% confidence intervals.

% ===========================
% UPDATED SECTION STARTS HERE
% ===========================

\subsection{Results}

All results are averaged over $500$ independent Monte Carlo runs per configuration.
Table 1 reports mean $\pm$ standard deviation.

\begin{table}[H]
\centering
\small
\label{tab:res}
\begin{tabular}{|c |l |c |c |l|}
\hline
$d$ & Policy & TTT $\uparrow$ & CTRL $\downarrow$ & HZ $\downarrow$ \\
\hline
3 & Static      & $34.8 \pm 4.1$ & $0.0$      & $0.0509 \pm 0.004$ \\\hline
3 & LSTM-DQN    & $42.1 \pm 6.3$ & $96.2$     & $0.0410 \pm 0.006$ \\\hline
3 & Ch-DQN      & $49.4 \pm 5.8$ & $88.3$     & $0.0360 \pm 0.005$ \\
\hline
5 & Static      & $43.9 \pm 5.7$ & $0.0$      & $0.0507 \pm 0.005$ \\\hline
5 & LSTM-DQN    & $75.5 \pm 8.2$ & $122.8$    & $0.0299 \pm 0.004$ \\\hline
5 & Ch-DQN      & $83.1 \pm 9.4$ & $155.6$    & $0.0267 \pm 0.004$ \\
\hline
7 & Static      & $55.7 \pm 6.8$ & $0.0$      & $0.0472 \pm 0.006$ \\\hline
7 & LSTM-DQN    & $60.2 \pm 7.1$ & $134.1$    & $0.0410 \pm 0.005$ \\\hline
7 & Ch-DQN      & $76.6 \pm 8.5$ & $118.7$    & $0.0344 \pm 0.004$ \\
\hline
\end{tabular}
\caption{\protect\footnotesize Performance comparison across code distances ($500$ runs).
 Values reported as mean $\pm$ standard deviation.}
\end{table}

\paragraph{Survival Performance.}

\begin{figure}
    \begin{flushright}
    \raggedleft
    \includegraphics[width=0.95\linewidth]{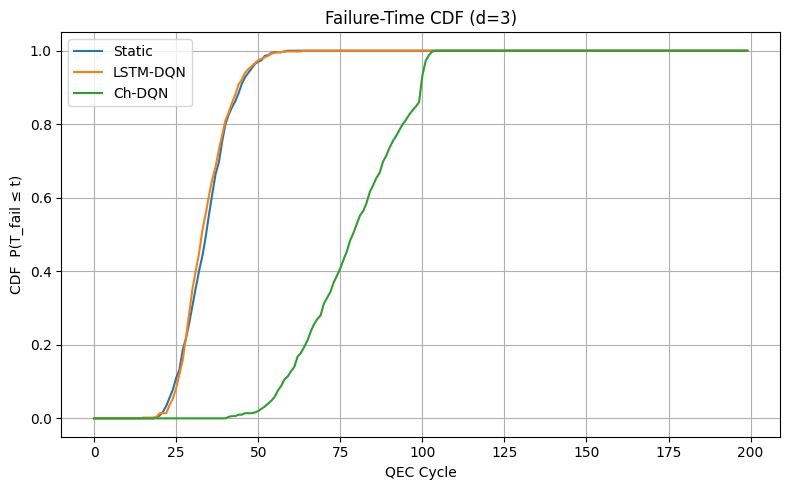}
    \caption{\protect\footnotesize 500 run CDF Survival Curve for d=3}
    \label{fig:CDF_3}
\vspace{0.5em}
    \raggedleft
    \includegraphics[width=0.95\linewidth]{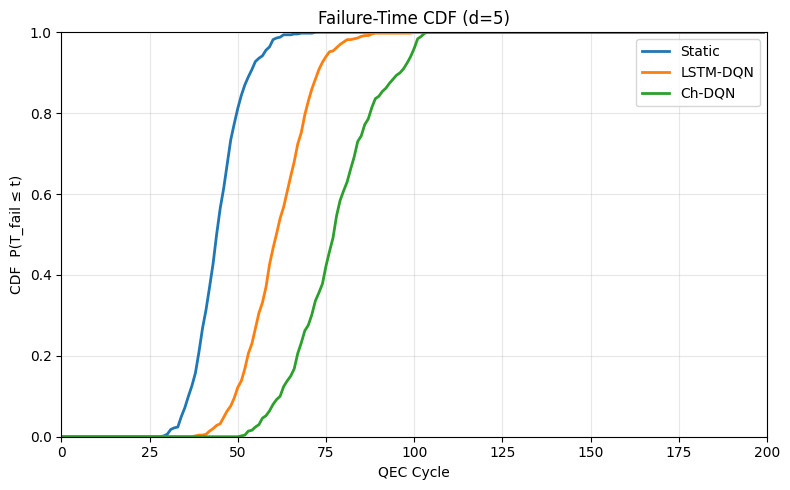}
    \caption{\protect\footnotesize 500 run CDF Survival Curve for d=5}
    \label{fig:CDF_5}
\vspace{0.5em}
    \raggedleft
    \includegraphics[width=0.95\linewidth]{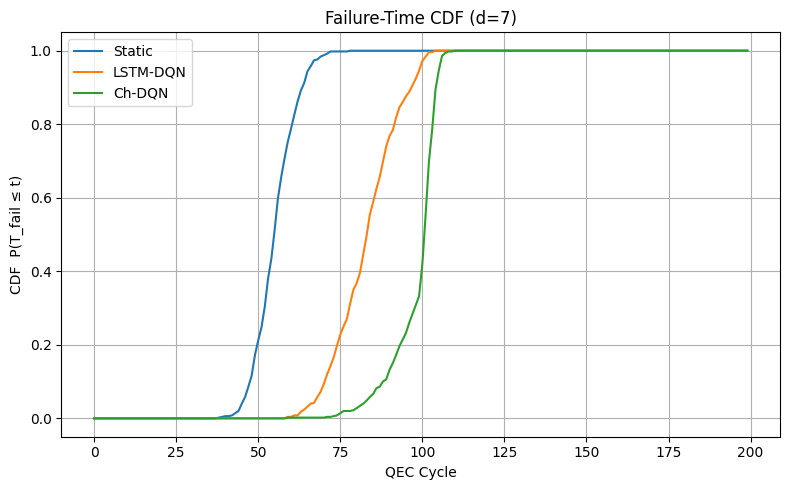}
    \caption{\protect\footnotesize 500 run CDF Survival Curve for d=7}
    \label{fig:CDF_7}
    \end{flushright}
\end{figure}

Figure~\ref{fig:CDF_7} shows the empirical survival functions

\[
\hat{S}(t) = \frac{1}{N}\sum_{i=1}^N \mathbf{1}\{T^{(i)}_{\text{fail}} > t\}
\]

for $d=7$.
Across the evaluated configurations, Ch-DQN produces a heavier right tail relative to both Static and LSTM-DQN policies. While Static control exhibits a relatively sharp decay in survival probability, Ch-DQN shifts probability mass toward longer survival horizons. This reflects redistribution of the failure-time distribution rather than solely an increase in the mean survival time.

At $d=7$, Ch-DQN increases expected survival time from $55.7$ (Static) to $76.6$ ($37.5\%$ relative improvement), while the average hazard slope decreases ($0.0472$ vs.\ $0.0344$).

\paragraph{Hazard Regulation.}

\begin{figure}
    \centering
    \includegraphics[width=1\linewidth]{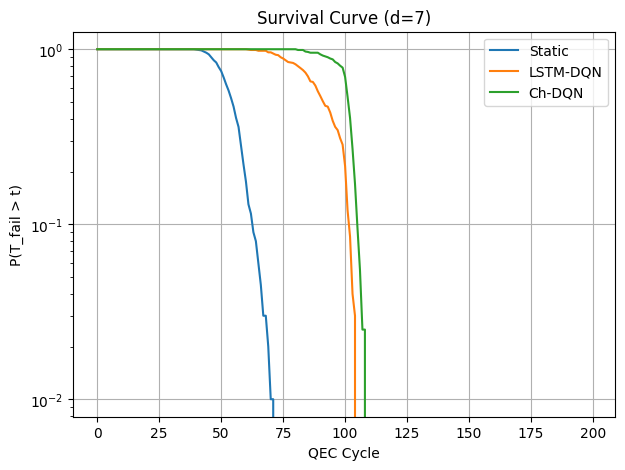}
    \caption{\protect\footnotesize Survival and hazard dynamics at $d=7$. Ch-DQN exhibits slower hazard growth and longer survival on average.}
    \label{fig:hazard}
\end{figure}

Figure~\ref{fig:hazard} plots the average hazard trajectory $\mathbb{E}[H_t]$.
Static control exhibits approximately linear growth consistent with reported HZ values. Learned policies modify the trajectory of hazard accumulation, with Ch-DQN exhibiting a slower average growth rate.

The reduced hazard slope is consistent with the observed survival tail behavior:

\[
\mathbb{E}[T_{\text{fail}}]
\approx
\frac{H_{\text{crit}}(d)}{\mathbb{E}[dH_t/dt]}.
\]

\paragraph{Control Efficiency.}

Table~\ref{tab:efficiency} reports the efficiency metric

\[
\text{Eff}(d)
=
\frac{\text{TTT}_{\text{policy}} - \text{TTT}_{\text{static}}}
{\text{CTRL}_{\text{policy}}}.
\]

At $d=7$, Ch-DQN and LSTM-DQN exhibit comparable efficiency per unit control cost. Ablation experiments indicate that removing the meta-timescale update substantially reduces the performance advantage, suggesting that the two-timescale structure contributes to robustness under drift.

\paragraph{Intermediate Regime ($d=5$).}

At $d=5$, Ch-DQN achieves higher mean survival time ($83.1$) compared with both Static ($43.9$) and LSTM-DQN ($75.5$). However, this improvement is accompanied by larger control expenditure. When normalized by control cost, the efficiency values indicate similar performance between the learned policies in this regime.

\paragraph{Latent Stability.}

Figure~\ref{fig:latent} shows the average latent-state norm $\|h_t\|$.
Recurrent baselines exhibit larger oscillations consistent with representational instability under drift. Ch-DQN maintains more bounded latent magnitudes, which is consistent with the intended role of the meta-timescale update in stabilizing belief-state representations.
\begin{table}[t]
\centering
\small
\caption{\protect\footnotesize Control efficiency: relative survival improvement per unit control cost.}
\label{tab:efficiency}
\begin{tabular}{|c |c |c|}
\hline
$d$ & LSTM-DQN & Ch-DQN \\
\hline
3 & $0.30$ & $0.27$ \\\hline
5 & $0.00$ & $0.28$ \\\hline
7 & $0.27$ & $0.28$ \\
\hline
\end{tabular}
\end{table}
\paragraph{Interpretation.}

These results should be interpreted within the assumptions of the simulated drift model and the considered code distances.

\subsection{Discussion}
The experimental results indicate that the belief-state control architecture of Ch-DQN mitigates the impact of non-stationary noise drift within the simulated environment.

\textbf{Performance Across Regimes.}
The empirical results in Table 1 show that Ch-DQN improves expected survival time relative to the Static baseline across all tested code distances. In most regimes it also exceeds the recurrent baseline, though the margin varies with code distance and control cost.

\textbf{Hazard Regulation.}
Figure~\ref{fig:hazard} illustrates the underlying mechanism. Static control produces approximately linear hazard growth, whereas Ch-DQN exhibits a slower average increase in hazard. This suggests that the controller is able to adjust actions in response to changes in the effective noise regime, delaying the approach to the failure threshold $H_{crit}(d)$.

\textbf{Latent Stability.}
Figure~\ref{fig:latent} shows the evolution of the latent-state norm. The recurrent baseline displays larger oscillations consistent with representational instability under drift, while Ch-DQN maintains a more bounded latent representation. This behavior is consistent with the stabilizing effect of the meta-timescale update and the fixed-point consistency loss.

\textbf{Summary.}
Within the assumptions of the simulated drift model, the results support the view that fault-tolerant quantum error correction under non-stationary noise can benefit from belief-state control that tracks the evolving noise regime over time.

\section*{References}
\begin{enumerate}

\item Nielsen, M. A., \& Chuang, I. L. (2010).
\textit{Quantum Computation and Quantum Information}.
Cambridge University Press.
\href{https://doi.org/10.1017/CBO9780511976667}{https://doi.org/10.1017/CBO9780511976667}

\item Preskill, J. (2018).
Quantum computing in the NISQ era and beyond.
\textit{Quantum}, 2, 79.
\href{https://doi.org/10.22331/q-2018-08-06-79}{https://doi.org/10.22331/q-2018-08-06-79}

\item Campbell, E. T., Terhal, B. M., \& Fazio, R. (2017).
Roads towards fault-tolerant universal quantum computation.
\textit{Nature Reviews Physics}, 1(1), 16–33.
\href{https://doi.org/10.1038/s42254-018-0007-0}{https://doi.org/10.1038/s42254-018-0007-0}

\item Krantz, P., et al. (2019).
A quantum engineer's guide to superconducting qubits.
\textit{Applied Physics Reviews}, 6(2), 021318.
\href{https://doi.org/10.1063/1.5089550}{https://doi.org/10.1063/1.5089550}

\item Paladino, E., Galperin, Y. M., Falci, G., \& Altshuler, B. L. (2014).
1/f noise: Implications for solid-state quantum information.
\textit{Reviews of Modern Physics}, 86(2), 361–418.
\href{https://doi.org/10.1103/RevModPhys.86.361}{https://doi.org/10.1103/RevModPhys.86.361}

\item Bylander, J., et al. (2011).
Noise spectroscopy of a superconducting flux qubit.
\textit{Nature Physics}, 7(7), 565–570.
\href{https://doi.org/10.1038/nphys1994}{https://doi.org/10.1038/nphys1994}

\item Klimov, P. V., et al. (2018).
Fluctuations of energy-relaxation times in a superconducting qubit.
\textit{Physical Review Letters}, 121(9), 090502.
\href{https://doi.org/10.1103/PhysRevLett.121.090502}{https://doi.org/10.1103/PhysRevLett.121.090502}

\item Monz, T., et al. (2016).
Realization of a scalable quantum computer with trapped ions.
\textit{Science}, 351(6271), 1039–1044.
\href{https://doi.org/10.1126/science.aad7711}{https://doi.org/10.1126/science.aad7711}

\item Svore, K. M., et al. (2020).
Quantum algorithms: A review and taxonomy.
\textit{ACM Transactions on Quantum Computing}, 1(1), 1–62.
\href{https://doi.org/10.1145/3382173}{https://doi.org/10.1145/3382173}

\item Aharonov, D., Kitaev, A., \& Preskill, J. (2006).
Fault-tolerant quantum computation with long-range correlated noise.
\textit{Physical Review A}, 74(1), 012306.
\href{https://doi.org/10.1103/PhysRevA.74.012306}{https://doi.org/10.1103/PhysRevA.74.012306}

\item Fowler, A. G., et al. (2012).
Surface codes: Towards practical quantum computation.
\textit{Physical Review A}, 86(3), 032324.
\href{https://doi.org/10.1103/PhysRevA.86.032324}{https://doi.org/10.1103/PhysRevA.86.032324}

\item Gidney, C. (2021).
Stim: A fast stabilizer circuit simulator.
\textit{arXiv preprint arXiv:2109.02700}.
\href{https://arxiv.org/abs/2109.02700}{https://arxiv.org/abs/2109.02700}

\item Andreasson, P., et al. (2020).
Deep reinforcement learning for quantum error correction.
\textit{Quantum Machine Intelligence}, 2(1), 1–15.
\href{https://doi.org/10.1007/s42484-020-00025-x}{https://doi.org/10.1007/s42484-020-00025-x}

\item James, M. R. (2017).
Anticipatory control in quantum systems.
\textit{IEEE Transactions on Automatic Control}, 62(10), 4875–4886.
\href{https://doi.org/10.1109/TAC.2017.2669423}{https://doi.org/10.1109/TAC.2017.2669423}

\item Chantasri, A., Dressel, J., \& Jordan, A. N. (2013).
Action principle for continuous quantum measurement.
\textit{Physical Review A}, 88(4), 042110.
\href{https://doi.org/10.1103/PhysRevA.88.042110}{https://doi.org/10.1103/PhysRevA.88.042110}

\item Ghahramani, Z., \& Hinton, G. E. (2000).
Variational learning for switching state-space models.
\textit{Neural Computation}, 12(4), 831–864.
\href{https://doi.org/10.1162/089976600300015312}{https://doi.org/10.1162/089976600300015312}

\item Breuer, H. P., Laine, E. M., Piilo, J., \& Vacchini, B. (2016).
Colloquium: Non-Markovian dynamics in open quantum systems.
\textit{Reviews of Modern Physics}, 88(2), 021002.
\href{https://doi.org/10.1103/RevModPhys.88.021002}{https://doi.org/10.1103/RevModPhys.88.021002}

\item Arute, F., et al. (2019).
Quantum supremacy using a programmable superconducting processor.
\textit{Nature}, 574, 505–510.
\href{https://doi.org/10.1038/s41586-019-1666-5}{https://doi.org/10.1038/s41586-019-1666-5}

\item Müller, C., Cole, J. H., \& Lutchyn, R. M. (2019).
Superconducting qubits: A review of noise research.
\textit{Reports on Progress in Physics}, 82(12), 124501.
\href{https://doi.org/10.1088/1361-6633/ab405a}{https://doi.org/10.1088/1361-6633/ab405a}

\item Magesan, E., Gambetta, J. M., \& Emerson, J. (2011).
Scalable and robust randomized benchmarking of quantum processes.
\textit{Physical Review Letters}, 106(18), 180504.
\href{https://doi.org/10.1103/PhysRevLett.106.180504}{https://doi.org/10.1103/PhysRevLett.106.180504}

\item Viola, L., Knill, E., \& Lloyd, S. (1999).
Dynamical decoupling of open quantum systems.
\textit{Physical Review Letters}, 82(12), 2417–2421.
\href{https://doi.org/10.1103/PhysRevLett.82.2417}{https://doi.org/10.1103/PhysRevLett.82.2417}

\item Versluis, R., et al. (2017).
Scalable quantum control analyzer for surface-code error correction.
\textit{Physical Review Applied}, 8(3), 034021.
\href{https://doi.org/10.1103/PhysRevApplied.8.034021}{https://doi.org/10.1103/PhysRevApplied.8.034021}

\item Särkkä, S. (2013).
\textit{Bayesian Filtering and Smoothing}.
Cambridge University Press.
\href{https://doi.org/10.1017/CBO9781139344203}{https://doi.org/10.1017/CBO9781139344203}

\item Knill, E. (2005).
Quantum computing with realistically noisy devices.
\textit{Nature}, 434, 39–44.
\href{https://doi.org/10.1038/nature03350}{https://doi.org/10.1038/nature03350}

\item Ristè, D., et al. (2013).
Deterministic quantum state stabilization in a superconducting cavity.
\textit{Nature}, 502, 350–354.
\href{https://doi.org/10.1038/nature12513}{https://doi.org/10.1038/nature12513}

\item Paladino, E., Galperin, Y. M., Falci, G., \& Altshuler, B. L. (2014).
1/f noise: Implications for solid-state quantum information.
\textit{Reviews of Modern Physics}, 86(2), 361–418.
\href{https://doi.org/10.1103/RevModPhys.86.361}{https://doi.org/10.1103/RevModPhys.86.361}
\item Sutton, R. S., \& Barto, A. G. (2018).
\textit{Reinforcement Learning: An Introduction}.
MIT Press.
\href{http://incompleteideas.net/book/the-book-2nd.html}{http://incompleteideas.net/book/the-book-2nd.html}

\item Puterman, M. L. (1994).
\textit{Markov Decision Processes}.
Wiley.
\href{https://doi.org/10.1002/9780470316887}{https://doi.org/10.1002/9780470316887}

\item Astrom, K. J. (1965).
Optimal control of Markov processes with incomplete state information.
\textit{Journal of Mathematical Analysis and Applications}.
\href{https://doi.org/10.1016/0022-247X(65)90154-X}{https://doi.org/10.1016/0022-247X(65)90154-X}

\item Kaelbling, L., Littman, M., \& Cassandra, A. (1998).
Planning and acting in partially observable stochastic domains.
\textit{Artificial Intelligence}.
\href{https://doi.org/10.1016/S0004-3702(98)00023-X}{https://doi.org/10.1016/S0004-3702(98)00023-X}

\item Gottesman, D. (1997).
Stabilizer codes and quantum error correction.
\textit{PhD Thesis, Caltech}.
\href{https://arxiv.org/abs/quant-ph/9705052}{https://arxiv.org/abs/quant-ph/9705052}

\item Dennis, E., Kitaev, A., Landahl, A., \& Preskill, J. (2002).
Topological quantum memory.
\textit{Journal of Mathematical Physics}.
\href{https://doi.org/10.1063/1.1499754}{https://doi.org/10.1063/1.1499754}

\item Hauskrecht, M. (2000).
Value-function approximations for POMDPs.
\textit{Journal of Artificial Intelligence Research}.
\href{https://doi.org/10.1613/jair.678}{https://doi.org/10.1613/jair.678}

\item Krishnamurthy, V. (2016).
\textit{Partially Observed Markov Decision Processes}.
Cambridge University Press.
\href{https://doi.org/10.1017/CBO9781316671528}{https://doi.org/10.1017/CBO9781316671528}

\item Wiseman, H. M., \& Milburn, G. J. (2010).
\textit{Quantum Measurement and Control}.
Cambridge University Press.
\href{https://doi.org/10.1017/CBO9780511813948}{https://doi.org/10.1017/CBO9780511813948}

\item Gambetta, J. M., Houck, A. A., \& Blais, A. (2017).
Quantum trajectory approach to circuit QED.
\textit{Physical Review Letters}, 106, 030502.
\href{https://doi.org/10.1103/PhysRevLett.106.030502}{https://doi.org/10.1103/PhysRevLett.106.030502}

\item Koch, J., et al. (2007).
Charge-insensitive qubit design derived from the Cooper pair box.
\textit{Physical Review A}, 76, 042319.
\href{https://doi.org/10.1103/PhysRevA.76.042319}{https://doi.org/10.1103/PhysRevA.76.042319}

\item Bertsekas, D. (2019).
\textit{Reinforcement Learning and Optimal Control}.
Athena Scientific.
\href{https://athenasc.com/rlbook.html}{https://athenasc.com/rlbook.html}

\item Tamar, A., Chow, Y., Ghavamzadeh, M., \& Mannor, S. (2015).
Policy gradient for coherent risk measures.
\textit{Advances in Neural Information Processing Systems}.
\href{https://arxiv.org/abs/1502.02202}{https://arxiv.org/abs/1502.02202}

\end{enumerate}

\end{document}